\documentclass[iop,apjl]{emulateapj}

\usepackage{amsmath}

\shorttitle{And XXVIII: A Distant Satellite of Andromeda}
\shortauthors{Slater, Bell, Martin}

\slugcomment{Accepted to ApJL, 21 Oct 2011.}

\begin{document}

\title{Andromeda XXVIII: A Dwarf Galaxy More Than 350 kpc from Andromeda}

\author{Colin T. Slater and Eric F. Bell}
\affil{Department of Astronomy, University of Michigan,
    500 Church St., Ann Arbor, MI 48109}

\and

\author{Nicolas F. Martin}
\affil{Max-Planck-Institut f\"{u}r Astronomie, K\"{o}nigstuhl 17,
    D-69117 Heidelberg, Germany}

\begin{abstract}
We report the discovery of a new dwarf galaxy, Andromeda XXVIII, using data from
the recently-released SDSS DR8. The galaxy is a likely satellite of Andromeda,
and, at a separation of $365^{+17}_{-1}$ kpc, would be one of the most distant of
Andromeda's satellites. Its heliocentric distance is $650^{+150}_{-80}$ kpc, and
analysis of its structure and luminosity show that it has an absolute magnitude
of $M_V = -8.5^{+0.4}_{-1.0}$ and half-light radius of $r_h = 210^{+60}_{-50}$
pc, similar to many other faint Local Group dwarfs. With presently-available
imaging we are unable to determine if there is ongoing or recent star formation,
which prevents us from classifying it as a dwarf spheroidal or dwarf irregular.
\end{abstract}


\section{Introduction}

In recent years the environment of Andromeda has been a prime location for the
discovery of dwarf galaxies and tidal structures, much of which has been enabled
by large surveys on the Isaac Newton Telescope \citep{ferguson02,irwin08} and
the Canada-France-Hawaii telescope \citep{ibata07, mcconnachie09, martin06,
martin09}. These surveys have obtained deep observations over a significant
fraction of the area within 180 kpc of Andromeda, and yielded a considerable
number of new discoveries.  In addition to these dedicated surveys, two
satellites of Andromeda have been found in the Sloan Digital Sky Survey (SDSS)
imaging \citep[And IX and X,][]{zucker04,zucker07}, using an early SDSS scan
targeting Andromeda specifically. More recently, the SDSS project has released
Data Release 8 \citep[DR8,][]{DR8}, which adds $\sim 2500$ deg$^2$ of imaging
coverage in the south Galactic cap and covers almost half of the area within
$35^\circ$ of Andromeda. While the SDSS is substantially shallower than the
dedicated M31 surveys, it is deep enough to enable the discovery of relatively
bright (by today's standards) dwarf galaxies.

It is in this new SDSS coverage that we report the discovery of a dwarf galaxy,
which we are preliminarily calling Andromeda XXVIII.  The dwarf is separated
from Andromeda by $27.7^\circ$ on the sky, which gives it a minimum distance to
M31 of $365$ kpc. This distance is significantly larger than the virial radius
of Andromeda \citep[$r_{\rm vir}=300$ kpc,][]{klypin02}. And XXVIII is
therefore one of a handful of known examples of dwarf galaxies that are less
likely to be significantly influenced by the environment of their host galaxy,
which makes them important test cases for theories of dwarf galaxy formation and
evolution.

\section{Detection}

At the distance of Andromeda \citep[$785 \pm 25$ kpc,][]{mcconnachie05},
searches for dwarf galaxies in the SDSS are limited to using red giant branch
(RGB) stars as tracers of the underlying population of main-sequence and
subgiant stars. Alternative tracers commonly used for detecting dwarf galaxies
around the Milky Way, such as horizontal branch or main sequence turn-off stars,
are much too faint to be detected.  To detect dwarf galaxies in SDSS we compute
star counts in $2^\prime \times 2^\prime$ bins, selecting only stars with $0.3 <
{\rm r-i} < 0.8$, colors roughly similar to metal-poor giant branch stars.
Overdensities are readily apparent upon visual inspection of the resulting map
as ``hot pixels'', typically with counts of 10-15 objects as compared to the
background of 1-3 objects per bin. Most of these overdensities are galaxy
clusters at intermediate redshift, which contain many spatially-unresolved
member galaxies that are erroneously classified as stars and have similar colors
as giant branch stars. Visual inspection of the SDSS image along with the
color-magnitude diagram is sufficient to reject these false-positives.

\begin{figure}
\epsscale{.90}
\plotone{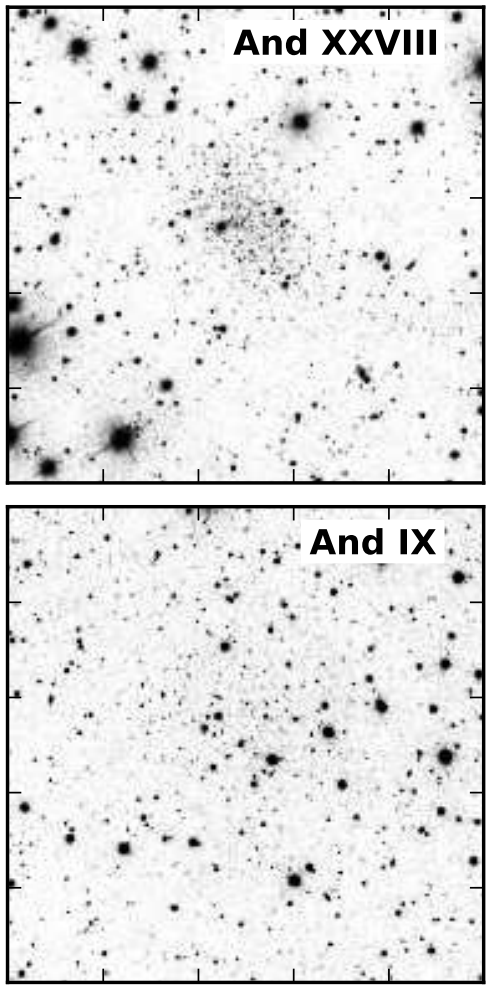}
\caption{SDSS image of And XXVIII ({\it top}), and, for comparison, an SDSS
image of And IX ({\it bottom}), which was also discovered in SDSS
\citep{zucker04}. Both images were obtained from the SDSS SkyServer, and are
$6.6^\prime \times 6.6^\prime$.  North is up, and east is to the left.
\label{sdss_image}} \end{figure}

\begin{figure*}
\epsscale{1.0}
\plotone{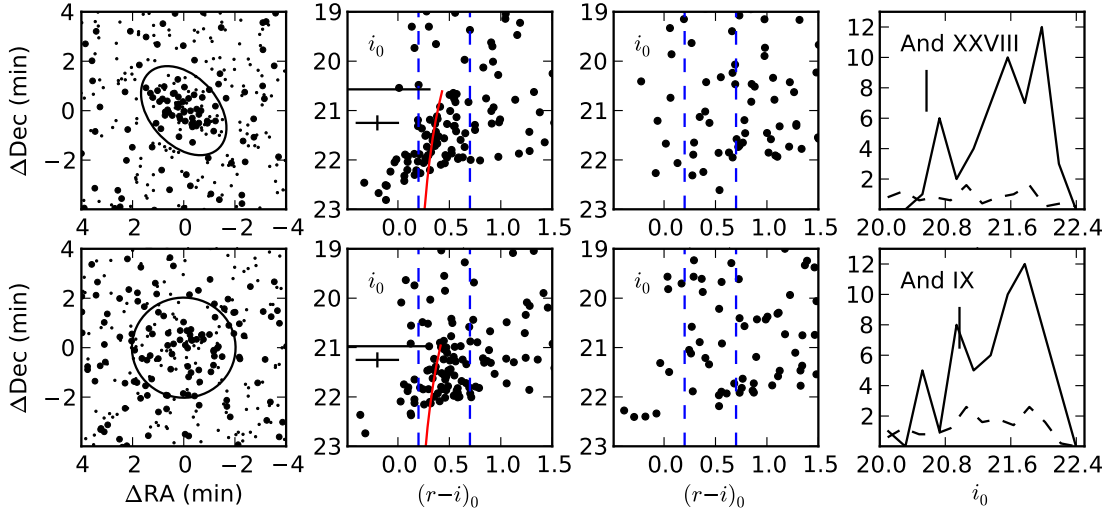}
\caption{Detection plots for And XXVIII ({\it top row}), with the same plots for
And IX shown for comparison ($M_V \sim -8.3$, {\it bottom row}). {\it Far left}:
the position of stars detected in SDSS is plotted, with stars that fall within
our color-cut as large points, and other stars as small points. An ellipse at
1.5 times the half light radius is also shown. {\it Middle left}:
color-magnitude diagram of stars inside twice the half-light radius. The
color-cut used to detect RGB stars is shown by the dashed vertical lines. An
isochrone from \citet{dotter08}  is overplotted ([Fe/H]=-2.0 for And XXVIII,
[Fe/H]=-2.2 for And IX), along with a horizontal line indicating the tip of the
red giant branch, and a representative photometric error bar on the left. {\it
Middle right}: color-magnitude diagram of a background annulus.  {\it Far
right}: luminosity function of the color-selected red giant stars (solid line),
and the background annulus (dashed).\label{cmd}} \end{figure*}

The SDSS image of And XXVIII is shown in Figure~\ref{sdss_image}, along with an
image of And IX for comparison, and the properties of And XXVIII are summarized
in Table~\ref{properties_table}. The color-magnitude diagram of the dwarf is
shown in Figure~\ref{cmd}, along with a CMD of the field region surrounding the
dwarf, a plot of measured star positions, and a histogram as a function of
i-band magnitude. These plots are also shown for And IX, another dwarf galaxy
that was discovered in SDSS. An isochrone from \citet{dotter08} of an old,
metal-poor system ($12$ Gyr old, [Fe/H] = -2.0) is also shown on the CMD to
illustrate the position of the red giant branch. An overdensity at $0.3 < ({\rm
r-i})_0 < 0.8$ is clearly visible. The RGB is very wide in color, owing to
considerable photometric uncertainty at very faint magnitudes in SDSS, which is
illustrated by the error bar on the left side of the CMD (estimated from repeat
imaging of SDSS stripe 82; \citealp{bramich08}).

\begin{table}
\caption{Properties of And XXVIII \label{properties_table}}
\begin{center}
\begin{tabular}{l c}
\tableline
Parameter & \\
\tableline 
\tableline
$\alpha$ (J2000) & $22^{\rm h}$ $32^{\rm m}$ $41\fs 2$  \\
$\delta$ (J2000) & $31\arcdeg$ $12\arcmin$ $58.2\arcsec$ \\
E(B-V) & 0.087\tablenotemark{a} \\
Ellipticity & $0.34$ $\pm$ $0.13$ \\
Position Angle (N to E) & $39^\circ$ $\pm$ $16$ \\
$r_h$ & $1\farcm 11$ $\pm$ $0\farcm 21$ \\
$r_h$ & $210^{+60}_{-50}$ pc \\
$D$ & $650^{+150}_{-80}$ kpc \\
$(m - M)_0$ & $24.1^{+0.5}_{-0.2}$ \\
$r_{\rm M31}$ & $365^{+17}_{-1}$ kpc\tablenotemark{b} \\
$M_V$ & $-8.5^{+0.4}_{-1.0}$ \\
\tableline
\tablenotetext{1}{\citet{SFD98}}
\tablenotetext{2}{Since the measured distance puts And XXVIII very close
to the tangent point along its line of sight, the uncertainty in $r_{\rm M31}$ is
very asymmetric.}
\end{tabular}
\end{center}
\end{table}

\section{Properties of And XXVIII}


We computed the distance to And XXVIII by measuring the magnitude of the tip of
the red giant branch (TRGB), which has a roughly constant absolute magnitude in
metal-poor stellar systems \citep{bellazzini01}. This method has been used
extensively for dwarf galaxies \citep[e.g.,][]{mcconnachie05,martin09}, since
the TRGB is often the only distinguishable feature in the color-magnitude
diagram of distant systems. 

\begin{figure}
\epsscale{.90}
\plotone{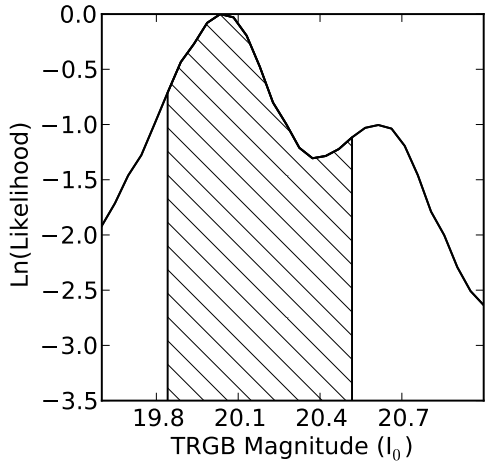}
\caption{Likelihood function of the TRGB position of And XXVIII, arbitrarily
normalized. The hatched region is the 67\% confidence interval. The secondary
maximum is clearly visible and less significant than the primary peak, but
cannot be ruled out. 
\label{trgb_likelihood}} 
\end{figure}

Quantitatively measuring the position of the TRGB is more complicated than it
would appear from looking at the color-magnitude diagram. This is especially
true in dwarf galaxies, where the giant branch is sparsely populated and the
small number counts lead to significant ``shot noise'' \citep{martin08}. We used
the maximum-likelihood estimator described in \citet{makarov06}, which modeled
the TRGB luminosity function as
\begin{equation}
\psi = \begin{cases}
10^{a(m - m_{\rm TRGB}) + b} & m - m_{\rm TRGB} \ge 0, \\
10^{c(m - m_{\rm TRGB})} & m - m_{\rm TRGB} < 0.
\end{cases}
\end{equation}
This broken power-law form takes three parameters: $a$ and $c$ are the slopes of
the luminosity function fainter and brigher than the TRGB, while $b$ is the
strength of the transition at the TRGB. We adopted the values from
\citet{makarov06} of $a = 0.3$ and $c=0.2$, and $b=0.6$. For the TRGB fit
we selected stars in our RGB color cuts with magnitudes $19.5 < i < 21.7$ to
avoid incompleteness at faint magnitudes. Though the data at the faintest
magnitudes are not critical for finding the position of breaks in the luminosity
function that might correspond to the TRGB, the faint end of the luminosity
function does affect our ability to determine the statistical significance of a
measured TRGB position. As a result we try to use as deep of
data as possible without reaching significant photometric incompleteness.  The
SDSS photometry was converted to Johnson I-band using the prescriptions of
\citet{jordi06}, and an intrinsic I-band magnitude of the TRGB was assumed of
$-4.04 \pm 0.12$ \citep{bellazzini01}. The likelihood function of the model as a
function of TRGB position is shown in Figure~\ref{trgb_likelihood}. We find that
the likelihood is maximized at $m_{I,{\rm TRGB}} = 20.1$, but a second peak also
appears at $m_{I,{\rm TRGB}}=20.6$ (in the Gunn-i filter, $20.6$ and $21.1$,
respectively).  This is the result of a clump of stars slightly fainter than
$m_I=20.1$, which causes the TRGB magnitude to change significantly depending on
whether or not they are included as part of the RGB. Though the fainter peak
cannot be ruled out, the TRGB magnitude we quote of $m_{I,{\rm TRGB}} =
20.1^{+0.5}_{-0.1}$ is the center of the more likely peak. The uncertainty on
this TRGB value is the 67\% confidence interval, which was computed by creating
a cumulative probability distribution function and measuring the 16.5\% through
83.5\% region. The resulting uncertainties are asymmetric, and this asymmetry
will propagate into all derived quantities, but this is a natural result of the
bimodal likelihood function.  The measured TRGB position yields a distance
modulus of $24.1^{+0.5}_{-0.2}$, which places the dwarf at a heliocentric
distance of $650^{+150}_{-80}$ kpc. Because this is very similar to the point of
closest approach to Andromeda along this line of sight (the ``tangent point''),
the distance between And XXVIII and M31 is largely insensitive to errors in the
heliocentric distance, and is measured to be $r_{\rm M31} = 365^{+17}_{-1}$ kpc. 

To measure the luminosity of And XXVIII, we computed luminosity functions from
SDSS data for three similar dwarf galaxies with known distances and luminosities
(And III, $M_V=-9.87 \pm 0.3$, \citealt{mcconnachie06}; And V, $M_V=-9.22 \pm
0.3$, \citealt{mcconnachie06}; And X, $M_V = -8.13 \pm 0.5$,
\citealt{zucker07}). We scaled these galaxies to a fiducial luminosity and
distance by correcting the dereddened apparent magnitude of each galaxy's stars
for their respective distances, and by scaling the number of stars in each
luminosity bin by the total luminosity of the galaxy. We then took the mean of
these profiles to produce a composite luminosity function that was less affected
by the ``shot noise'' inherent in such low number count systems. Since our
comparison objects span a range of distances, we applied a faint-magnitude cut
to ensure that the luminosity function of even the most distant comparison dwarf
was still photometrically complete. For stars with colors typical of RGB stars
we find that the data are complete to approximately $i = 21.7$ (non-dereddened).
Since the most distant comparison dwarf has a distance modulus that is greater
than than of And XXVIII by 0.4, our corresponding completeness cut on And XXVIII
for the luminosity function comparison was $i=21.3$ ($i_0 = 21.5$).  This
binned, composite luminosity function was then scaled to match that of And
XXVIII (again using a maximum likelihood method to properly account for
Poissonian uncertainties, and with uncertainties on the comparison dwarfs'
luminosities included), and the scaling factor thus determined the luminosity of
the galaxy relative to the fiducial luminosity. This method produces results
largely equivalent to the method of \citet{martin08} for relatively bright
dwarfs. The luminosity determined by this method is $M_V = -8.5^{+0.4}_{-1.0}$
(the large uncertainty is primarily due to the uncertainity in the distance
measurement) is generally similar to that of other local group dwarfs. To ensure
that issues of photometric completeness or other systematics did not bias our
composite luminosity function, we also constructed a luminosity function from
deep observations of the Draco dwarf \citep[obtained on the Canada-France-Hawaii
Telescope,][]{segall07}, and used the same scaling method to measure the
luminosity of And XXVIII, which resulted in an identical value.  As a final
check, we compared And XXVIII to the model luminosity functions of
\citet{dotter08}, and again obtained a luminosity that is in good agreement with
the other methods ($M_V = -8.32$). 


The considerable scatter in color of the RGB stars due to photometric error
makes it difficult to determine the metallicity of the galaxy. This uncertainty
is illustrated by the CMD of And IX (Figure~\ref{cmd}, bottom middle-left),
which was measured by \citet{collins10} with deep imaging to have [Fe/H] =
$-2.2$ $\pm$ $0.2$. Though the fit to the Dotter isochrone is very good in the
\citet{collins10} data, the SDSS data show significant scatter in color and 
appear to be systematically offset in color from the isochrone. It is unclear
whether this is the result of inaccuracies in the isochrone or calibration error
at very faint magnitudes in the SDSS, but because of these uncertainties, it is
not be possible to constrain the metallicity of the galaxy with the observations
available. We can nevertheless say that the CMD of And XXVIII is not obviously
dissimilar to other metal-poor dwarf galaxies.

\begin{figure}
\epsscale{1.1}
\plotone{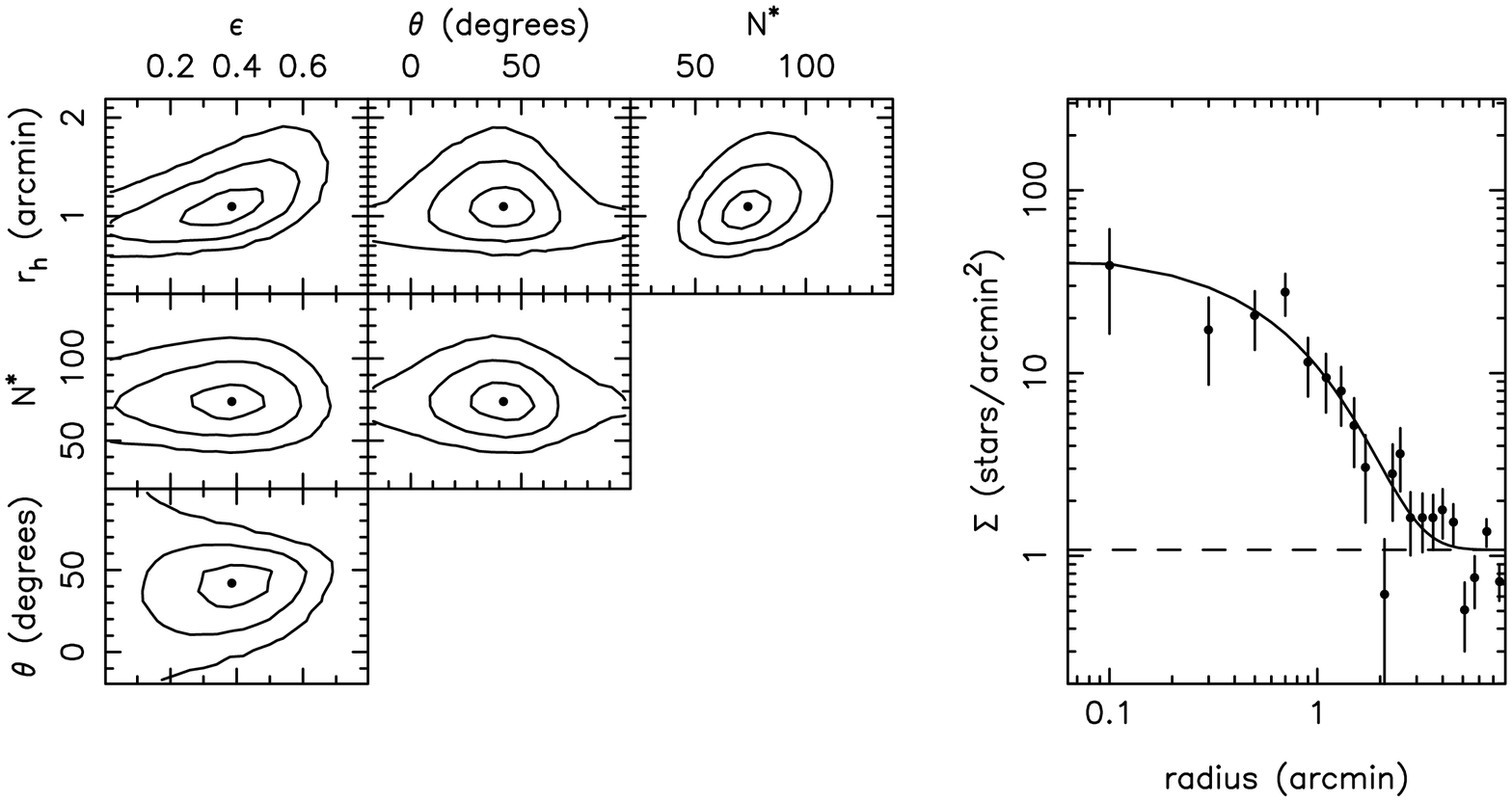}
\caption{{\it Left:} Confidence areas for the measurement of half-light radius,
ellipticity, position angle, and number of detected stars. The contours
correspond, when projected on the axes, to 1-, 2-, and 3-$\sigma$ uncertainties
(to allow reading of the marginalized $1-\sigma$ value straight from the plot
for each parameter). The filled circles correspond to the peak of the maximum
likelihood function. {\it Right:} Radial profile of And XXVIII, where stars have
been binned according to the best-fit structural parameters with Poisson
uncertainties on each bin. The solid black line is the best-fit exponential
profile, while the dashed horizontal line is the measured background level.
\label{profile_structure}} 
\end{figure}

We computed the radial profile of And XXVIII, along with the position, half-light
radius, eccentricity, and position angle using the maximum likelihood technique
described by \citet{martin08}. This method assumes an exponential profile for
the dwarf galaxy and a constant background level. Figure~\ref{profile_structure}
shows on the left maximum likelihood contours of the half-light radius,
ellipticity ($\epsilon$), position angle ($\theta$), and number of detected
stars in the overdensity within the SDSS data ($N_\star$), while the right side
shows the radial profile fit. The structural parameters have one-dimensional
1-~, 2-~, and 3-$\sigma$ confidence areas overlaid.  And XXVIII is
well-populated enough, even in the relatively shallow SDSS data, to permit easy
determination of these parameters without large uncertainties. The fact that
$N_\star = 0$ is excluded at $\gg3\sigma$ provides a quantitative indication
that this overdensity is unlikely to be a statistical artifact. The fact that
the half-light radius is well-determined also gives confidence that the
overdensity is real, since the fitting procedure usually finds unreasonably
large values for $r_h$ when run on non-galaxies. The half-light radius of $r_h =
210^{+60}_{-50}$ pc is typical of other Local Group dwarf galaxies and is
roughly the size of Draco. The position angle has a considerable uncertainty
associated with it, along with some covariance with ellipticity. These factors
may make the ellipse in the top-left panel of Figure~\ref{cmd} appear slightly
misaligned when juding the fit by eye. 

\section{Discussion}

Throughout this work we have referred to the newly discovered dwarf galaxy as
Andromeda XXVIII, but this may not be the most accurate identifier to use. The
dwarf is actually located in the constellation Pegasus, and could also be
identified as Pegasus II, as is the convention with Milky Way and Local Group
satellites.  However, its properties make it a likely satellite of M31, and
hence we follow the convention of naming satellites of M31 with the prefix
Andromeda regardless of their actual position. Since we have neither its radial
velocity nor its proper motion, we certainly cannot say whether the dwarf galaxy
is bound to M31,  but its distance to M31 is within the range of other galaxies
in the M31 system, and it is much further from the Milky Way than we would
expect for dwarfs bound to the Milky Way. If, on further study, the galaxy is
determined to be unbound from M31, then it should properly be referred to as
Pegasus II.  Further discussion of the complexities of dwarf galaxy names can be
found in the Appendix of \citet{martin09}.

The most intriguing feature of And XXVIII is its large distance from Andromeda,
which suggests that it might not have been strongly affected by interactions
with other galaxies. This could make it a prime test case for studies of dwarf
galaxy formation. The morphology and star formation history of And XXVIII are of
particular interest, as dwarf galaxies in the Local group that lay beyond
300-400 kpc from their host tend to be dwarf irregular galaxies, while those in
close proximity with their host tend to be dwarf spheroidals.  This
morphology-density relationship \citep{grebel03} is not without exceptions; for
instance, the dwarf spheroidals Tucana, Cetus, and the possible dwarf spheroidal
And XVIII are all more than $400$ kpc from the nearest non-dwarf galaxy.  These
distant dwarf spheroidals are a unique test for theories which suggest that
dwarf spheroidals form from dwarf irregulars via tidal interactions or ram
pressure stripping \citep{mayer06,weisz11}, since these galaxies could be
in the beginning stages of such a process and could exhibit evidence of such an
ongoing transformation. If And XXVIII were confirmed to be a dwarf spheroidal
without any recent star formation, it would add another test case for these
theories.

Alternatively, if star formation is detected in And XXVIII, it would be one of the
lowest mass star-forming galaxies known, and roughly analagous to LGS 3
\citep{thuan79} or Leo T \citep{irwin07}. The ability of such low mass galaxies
to retain gas and form stars is poorly understood, and identifying another
member of this class of galaxies would be a benefit to efforts to further
elucidate their nature.

Unfortunately, with shallow SDSS imaging we cannot conclusively determine
whether or not And XXVIII has ongoing or recent star formation. From the CMD of
Leo~T, the blue-loop stars that indicate recent star formation
are roughly 1.5 - 2 magnitudes fainter than the tip of the red giant branch.
Since the SDSS data of And XXVIII only extend approximately one magnitude below
the TRGB, blue-loop stars are not detectable.  We have also looked for H{\sc I}
in the galaxy using the Leiden/Argentine/Bonn Survey \citep{kalberla05}. This
survey clearly detects the H{\sc I} gas present in LGS 3 and Leo T
\citep[$M_{\rm HI} = 1.6 \times 10^5$, $\sim 4.3 \times 10^5 M_\odot$,
respectively,][]{grcevich09}, but shows no emission from And XXVIII. This could,
however, be the result of the velocity of the dwarf falling outside the
bandwidth used for the survey ($-400 < v_{\rm LSR} < 400$ km s$^{-1}$), so a
conclusive determination of the H{\sc I} gas content will require a measurement
of the radial velocity of the galaxy.

Though the exact significance of And XXVIII will not be known until follow-up
observations are conducted, it is clear that dwarf galaxies in the outer regions
of the Local Group are in a unique environment that enables their detailed study
before their properties are significantly altered by interactions with their 
host galaxy upon infall. Increasing the sample of nearby but isolated dwarfs
thus provides the data necessary to advance theories of dwarf galaxy formation
and evolution.

\acknowledgments

This work was partially supported by NSF grant AST 1008342. NFM acknowledges
funding by Sonderforschungsbereich SFB 881 ``The Milky Way System'' (subproject
A3) of the German Research Foundation (DFG). We would like to thank the referees
for their careful comments which improved the work.

Funding for SDSS-III has been provided by the Alfred P. Sloan Foundation, the
Participating Institutions, the National Science Foundation, and the U.S.
Department of Energy.  The SDSS-III web site is http://www.sdss3.org/. 

SDSS-III is managed by the Astrophysical Research Consortium for the
Participating Institutions of the SDSS-III Collaboration including the
University of Arizona, the Brazilian Participation Group, Brookhaven National
Laboratory, University of Cambridge, University of Florida, the French
Participation Group, the German Participation Group, the Instituto de Astroﬁsica
de Canarias, the Michigan State/Notre Dame/JINA Participation Group, Johns
Hopkins University, Lawrence Berkeley National Laboratory, Max Planck Institute
for Astrophysics, New Mexico State University, New York University, the Ohio
State University, University of Portsmouth, Princeton University, University of
Tokyo, the University of Utah, Vanderbilt University, University of Virginia,
University of Washington, and Yale University. 

{\it Facilities:} \facility{Sloan}.

\clearpage
\end{document}